\documentclass[a4paper,12pt]{article}
\usepackage{jheppubmod}
\usepackage{amsmath}
\usepackage{amssymb}
\usepackage{bbm}
\usepackage{bm}
\usepackage{verbatim}
\usepackage{amsfonts}
\usepackage{tipa}
\usepackage{mathrsfs}
\usepackage{mathtools}

\usepackage{graphicx}

\title{Gauge Invariant Descriptions of Gluon Polarizations Revisited}

\author{Zhi-Qiang Guo}
\author{and Iv\'{a}n Schmidt}
\emailAdd{zhiqiang.guo@usm.cl}
\emailAdd{ivan.schmidt@usm.cl}
\affiliation{Departamento de F\'{i}sica y Centro Cient\'{i}fico
Tecnol\'{o}gico de Valpara\'{i}so,\\ Universidad T\'{e}cnica Federico
Santa Mar\'{i}a,\\ Casilla 110-V, Valpara\'{i}so, Chile}

\abstract{We examine the feasibility of gauge invariant descriptions of the gluon polarization following the proposal that a gauge field can be decomposed into its physical part and its pure gauge part. We show that gauge invariant angular momentum currents can be constructed from summations of gauge variant Noether currents. We present novel expressions of the pure gauge field, which are used to formulate gauge invariant descriptions of the gluon spin and the photon spin. We show that the gauge invariant extension of the Chern-Simons current can describe the spins of the Laguerre-Gauss laser modes. We also discuss the relation of gauge invariant operators and the parton distributions constructed from Dirac variables.}


\begin{document}
\maketitle

\section{Introduction}\label{sec:1}

The electromagnetic and strong interactions are well recognized as gauge theories. Nevertheless, compared to their elegant mathematical formulation, gauge symmetries play an perplexing role when we try to understand the angular momentum structure of the photon field and gluon field. On one hand, the spin and orbital angular momentum of light modes have been measured experimentally~\cite{Beth:1936zz,ONeil:2002z,Leach:2002zz}, and the gluon spin contributions to the nucleon spin are also believed to be measurable~\cite{Ageev:2005pq,Boyle:2006ab,Kiryluk:2005mh}. On the other hand, it is believed that a gauge invariant operator decomposition of the gluon spin and orbital angular momentum is not feasible, compared to a well established gauge invariant decomposition of the quark spin and orbital angular momentum~\cite{Ji:1996ek}.

Recently, Chen et al.~\cite{Chen:2008ag} proposed that the gluon spin and orbital angular momentum can be identified gauge-independently by decomposing the gauge field into its physical part $A^{\mu}_{\mathrm{phys}}$ and its pure gauge part $A^{\mu}_{\mathrm{pure}}$. The decomposition of Chen et al. shows the appealing feature that the gauge covariant derivative operator in their construction appears to depend solely on the pure gauge field, so it can be regarded as the closest akin to the interaction-free canonical ones. Moreover, Chen et al. argue that the Coulomb gauge construction plays an especially superior role, mainly because the Abelian gauge field can be uniquely decomposed into its transverse part and longitudinal one. The proposal of Chen et al. has inspired lots of considerations on the feasibility of gauge invariant operator descriptions of the gluon spin~\cite{Wakamatsu:2010cb,Chen:2011gn,Hatta:2011zs,Zhang:2011rn,Lorce:2012ce}. However, there are also some concerns about the decomposition of Chen et al., such as the frame-dependence problem in the Coulomb gauge case and the light-cone gauge case~\cite{Ji:2012gc}, the uniqueness problem of identifications of the gluon spin operator~\cite{Lorce:2012rr,Wakamatsu:2013voa}, whether the requirement of gauge invariant operators is necessary~\cite{Leader:2011za}, and whether the decomposition of Chen et al. can provide new understandings on the decomposition of the nucleon spin beyond the framework of gauge invariant extensions~(GIEs)~\cite{Ji:2012gc}. In this paper, we discuss the feasibility of gauge invariant descriptions of the gluon spin and attempt to shed light on these concerns from several different aspects.

Firstly, we propose that gauge invariant expressions of angular momentum current can be constructed from summations of two kinds of Noether currents: one kind of current is induced by the Lorentz transformation and the other one is induced by the gauge transformation. These results are presented in section \ref{sec:2}. Secondly, in section~\ref{sec:3} we propose novel expressions for the pure gauge field $A^{\mu}_{\mathrm{pure}}$, using operators similar to the operators of specified twist in the operator product expansions. These expressions are apparently frame-independent. Based on these expressions, we suggest gauge invariant decompositions of the nucleon spin in section~\ref{sec:4.1}. We then pay special attention on the photon spin operator in section~\ref{sec:4.2}. When restricted on the Abelian case, we show that our proposed photon spin operator can describe the spin of light modes with a fixed frequency, such as the Laguerre-Gaussian laser modes. We further discuss the parton distribution functions constructed from Dirac variables~\cite{Dirac:1955uv,Pervushin:2001kq,Chen:2012vg,Lorce:2013gxa} and their relations to the gauge invariant quark momentums in section~\ref{sec:4.3}. Finally, we give conclusions in section~\ref{sec:5}. We also have three appendices to give more details of the body of our paper.

\section{Gauge Invariant Angular Momentum Currents As Summations}\label{sec:2}

For gauge theories, the conserved currents constructed from the Noether theorem are generally not gauge invariant. However, we show in this section that gauge invariant currents can be constructed from the summations of gauge variant currents. We consider the Lagrangian
\begin{eqnarray}
\label{sec-2-qcd-lag}
\mathcal {L}&=&-\frac{1}{4}F^{a}_{\mu\nu}F^{a\mu\nu}
+\frac{i}{2}[\bar{\psi}\gamma^{\mu}(\partial_{\mu}-igA_{\mu})\psi-(\partial_{\mu}\bar{\psi}+ig\bar{\psi}A_{\mu})\gamma^{\mu}\psi].
\end{eqnarray}
For the Lorentz invariance of this Lagrangian, the corresponding Noether current can be written as
\begin{eqnarray}
\label{sec-2-noether-lt}
{\mathcal{J}}^{\mu}_{LT}=\delta{x}^{\mu}\mathcal {L}+\frac{\partial\mathcal {L}}{\partial(\partial_{\mu}A_{\nu})}\delta_{LT}{A}_{\nu}+\frac{\partial\mathcal {L}}{\partial(\partial_{\mu}\psi)}\delta_{LT}\psi+\delta_{LT}\bar{\psi}\frac{\partial\mathcal {L}}{\partial(\partial_{\mu}\bar{\psi})},
\end{eqnarray}
with field variations up to first order of the parameters $\omega_{\mu\nu} $ as
\begin{eqnarray}
\label{sec-2-noether-lt-gauge}
\delta_{LT}{A}_{\mu}(x)&=&-y^{\beta}\partial_{\beta}A_{\mu}(x)+\omega_{\mu\beta}A^{\beta}(x),
\hspace{2mm}y_{\mu}=\delta{x}_{\mu}=\omega_{\mu\nu}x^{\nu},\\
\label{sec-2-noether-lt-fermion}
\delta_{LT}{\psi}(x)&=&-\frac{i}{2}\omega_{\alpha\beta}\mathcal{S}^{\alpha\beta}{\psi}(x)-y^{\beta}\partial_{\beta}\psi(x),\\
\label{sec-2-noether-lt-fermion-anti}
\delta_{LT}{\bar{\psi}}(x)&=&\frac{i}{2}{\bar{\psi}}(x)\omega_{\alpha\beta}\mathcal{S}^{\alpha\beta}-y^{\beta}\partial_{\beta}{\bar{\psi}}(x).
\end{eqnarray}
When Eq.~(\ref{sec-2-noether-lt}) is written in a manifest way, it gives the canonical angular momentum current as proposed by Jaffe and Manohar~\cite{Jaffe:1989jz}. The Lagrangian (\ref{sec-2-qcd-lag}) is also invariant under gauge transformations. For gauge transformations, the corresponding Noether current is given by
\begin{eqnarray}
\label{sec-2-noether-gt}
{\mathcal{J}}^{\mu}_{GT}=\frac{\partial\mathcal {L}}{\partial(\partial_{\mu}A_{\nu})}\delta_{GT}{A}_{\nu}+\frac{\partial\mathcal {L}}{\partial(\partial_{\mu}\psi)}\delta_{GT}\psi+\delta_{GT}\bar{\psi}\frac{\partial\mathcal {L}}{\partial(\partial_{\mu}\bar{\psi})},
\end{eqnarray}
with field variations up to first order of the gauge parameter $\theta $ as
\begin{eqnarray}
\label{sec-2-noether-gt-gauge}
\delta_{GT}{A}_{\mu}(x)&=&D_{\mu}\theta(x)=\partial_{\mu}\theta(x)-ig[A_{\mu},\theta(x)],\\
\label{sec-2-noether-gt-fermion}
\delta_{GT}{\psi}(x)&=&ig\theta(x)\psi(x),\hspace{2mm}\delta_{GT}{\bar{\psi}}(x)=-ig{\bar{\psi}}(x)\theta(x).
\end{eqnarray}
Written manifestly, the current in Eq.~(\ref{sec-2-noether-gt}) is
\begin{eqnarray}
\label{sec-2-noether-gt-form}
{\mathcal{J}}^{\mu}_{GT}&=&-2\mathrm{Tr}(F^{\mu\nu}D_{\nu}\theta)-g\bar{\psi}\gamma^{\mu}\theta\psi,\\
\label{sec-2-noether-gt-form-1}
&=&2\mathrm{Tr}(D_{\nu}F^{\mu\nu}\theta)-g\bar{\psi}\gamma^{\mu}\theta\psi-2\partial_{\nu}\mathrm{Tr}(F^{\mu\nu}\theta).
\end{eqnarray}
Here and hereafter we use the normalization $\mathrm{Tr}(T^{a}T^{b})=\frac{1}{2}\delta^{ab}$. The first two terms of Eq.~(\ref{sec-2-noether-gt-form-1}) give zero by the equations of motion. The third term satisfies $\partial_{\mu}\partial_{\nu}\mathrm{Tr}(F^{\mu\nu}\theta)=0$. So the current ${\mathcal{J}}^{\mu}_{GT}$ is conserved for any values of the field $\theta(x)$. We can suppose a parametrization of the field $\theta(x)$ as
\begin{eqnarray}
\label{sec-2-theta}
\theta(x)=-\omega_{\alpha\beta}x^{\alpha}N^{\beta}(x).
\end{eqnarray}
Then the current ${\mathcal{J}}^{\mu}_{GT}$ can be summed with ${\mathcal{J}}^{\mu}_{LT}$ to formulate a new current
\begin{eqnarray}
\label{sec-2-noether-lt-gt}
{\mathcal{J}}^{\mu}&=&{\mathcal{J}}^{\mu}_{LT}+{\mathcal{J}}^{\mu}_{GT}=\frac{1}{2}\omega_{\alpha\beta}M^{\mu\alpha\beta},\\
\label{sec-2-noether-generator}
M^{\mu\alpha\beta}&=&M_{qs}^{\mu\alpha\beta}+M_{qo}^{\mu\alpha\beta}+M_{gs}^{\mu\alpha\beta}+M_{go}^{\mu\alpha\beta},
\end{eqnarray}
with identifications of quark parts as
\begin{eqnarray}
\label{sec-2-noether-quark-spin}
M_{qs}^{\mu\alpha\beta}&=&\frac{1}{2}\epsilon^{\mu\alpha\beta\rho}\bar{\psi}\gamma_{\rho}\gamma_{5}\psi,\\
\label{sec-2-noether-quark-orbit}
M_{qo}^{\mu\alpha\beta}&=&\frac{i}{2}[\bar{\psi}\gamma^{\mu}x^{\alpha}(\partial^{\beta}-igN^{\beta})\psi-(\alpha\longleftrightarrow
\beta)]+\mathrm{H.C.}\\
&+&({\eta}^{\mu\alpha}x^{\beta}-{\eta}^{\mu\beta}x^{\alpha}){\mathcal{L}}_{\mathrm{quark}},\nonumber
\end{eqnarray}
and identifications of gluon parts as
\begin{eqnarray}
\label{sec-2-noether-gluon-spin}
M_{gs}^{\mu\alpha\beta}&=&-2\mathrm{Tr}\left[F^{\mu\alpha}(A^{\beta}-N^{\beta})-F^{\mu\beta}(A^{\alpha}-N^{\alpha})\right],\\
\label{sec-2-noether-gluon-orbit}
M_{go}^{\mu\alpha\beta}&=&2\mathrm{Tr}\left[F^{\mu}_{\hspace{2mm}\nu}(F^{\nu\alpha}x^{\beta}-F^{\nu\beta}x^{\alpha})\right]\\
&+&2\mathrm{Tr}\left[x^{\beta}F^{\mu}_{\hspace{2mm}\nu}D^{\nu}(A^{\alpha}-N^{\alpha})-(\alpha\longleftrightarrow
\beta)\right]\nonumber\\
&+&({\eta}^{\mu\alpha}x^{\beta}-{\eta}^{\mu\beta}x^{\alpha}){\mathcal{L}}_{\mathrm{gluon}}.\nonumber
\end{eqnarray}
Here ${\mathcal{L}}_{\mathrm{quark}}$ and ${\mathcal{L}}_{\mathrm{gluon}}$ are respectively the corresponding gauge invariant fermion part and gluon part of the Lagrangian~(\ref{sec-2-qcd-lag}), and $D_{\mu}=\partial_{\mu}-ig[A_{\mu},\cdot\hspace{0.5mm}]$. The above results are identical with our previous ones obtained by using a generalized Lorentz transformation~\cite{Guo:2012wv}. The cause of their consistency is simply that the Lie algebra of the generalized Lorentz transformation used in~\cite{Guo:2012wv} is the direct sum of the algebra of the Lorentz transformation and that of the gauge transformation. For $N^{\beta}=A^{\beta}$, we obtain the gauge invariant decompositions of Ji~\cite{Ji:1996ek}; For $N^{\beta}=A^{\beta}_{\mathrm{pure}}$, we obtain the gauge invariant decompositions of Chen et al.~\cite{Chen:2008ag}. The feasibility of derivation of gauge invariant currents by the Noether procedure is also discussed in~\cite{Zhou:2011z,Lorce:2013gxa} through different methods. Their results correspond to special cases of our above discussions when $N^{\beta}=A^{\beta}_{\mathrm{pure}}$. Harindranath and Kundu~\cite{Harindranath:1998ve}, Shore and White~\cite{Shore:1999be} ever showed that the difference between the decomposition of Ji and that of Jaffe and Manohar is just a total divergence by imposing equations of motion. Our above results are also consistent with their analysis in the case $N^{\beta}=A^{\beta}$.

\section{Novel Expressions for the Pure Gauge Field}\label{sec:3}

\subsection{Abelian Case}\label{sec:3.1}

A key step of the proposal of Chen et al. is to express the pure gauge field $A^{\mu}_{\mathrm{pure}}$ in terms of $A^{\mu}$ in order that no extra freedoms are introduced. For several expressions of the pure gauge field, see~\cite{Chen:2011gn,Hatta:2011zs,Zhang:2011rn,Lorce:2012ce}. In this section, we attempt to derive expressions for $A^{\mu}_{\mathrm{pure}}$ through a method introduced by Delbourgo and Thompson~\cite{Delbourgo:1986wz}. For an Abelian gauge field, we consider the pure gauge field determined by
\begin{eqnarray}
\label{sec-3-1-pure-gauge-eq}
\partial^{\mu}[A_{\mu}(x)-\partial_{\mu}\varphi(x)]=0,
\end{eqnarray}
which can be rewritten as
\begin{eqnarray}
\label{sec-3-1-pure-gauge-eq-re}
\partial^{\mu}\partial_{\mu}\varphi(x)=\mathcal{F}(x),\hspace{2mm}\mathcal{F}(x)=\partial^{\mu}A_{\mu}(x).
\end{eqnarray}
A series solution based on the inverse derivative operator has been given in~\cite{Delbourgo:1986wz,Kubo:1986ps,Guo:2012wv}.  We try to find another kind of solution of Eq.~(\ref{sec-3-1-pure-gauge-eq-re}), we make the ansatz
\begin{eqnarray}
\label{sec-3-1-pure-gauge-ansatz}
\varphi(x)=\sum^{\infty}_{s=0}\sum^{[\frac{s}{2}]}_{n=0}a_{s,n}(x^{\rho}x_{\rho})^{n+1}
x^{\alpha_1}\cdots{x}^{\alpha_{s-2n}}\partial_{\alpha_1}\cdots\partial_{\alpha_{s-2n}}
(\partial^{\beta}\partial_{\beta})^{n}\mathcal{F}(x),
\end{eqnarray}
where the notation $[\frac{s}{2}]$ stands for the largest integer which is smaller than $\frac{s}{2}$, and $a_{s,n}$ are some unknown constants to be determined later on. In the ansatz~(\ref{sec-3-1-pure-gauge-ansatz}), $s+1$ counts the number of the derivatives appearing in the expansions. The terms in the ansatz~(\ref{sec-3-1-pure-gauge-ansatz}) look very similar to the operators of specified twist appearing in the conventional operator product expansions. The ansatz~(\ref{sec-3-1-pure-gauge-ansatz}) has the simple property, that is, after being acted by the derivative operator twice,  it is transformed to be
\begin{eqnarray}
\label{sec-3-1-pure-gauge-ansatz-der}
\partial^{\mu}\partial_{\mu}\varphi(x)=\sum^{\infty}_{s=0}\sum^{[\frac{s}{2}]}_{n=0}{c}_{s,n}(x^{\rho}x_{\rho})^{n}
x^{\alpha_1}\cdots{x}^{\alpha_{s-2n}}\partial_{\alpha_1}\cdots\partial_{\alpha_{s-2n}}
(\partial^{\beta}\partial_{\beta})^{n}\mathcal{F}(x).
\end{eqnarray}
It remains a similar structure to Eq.~(\ref{sec-3-1-pure-gauge-ansatz}), but its coefficients are modified to be
\begin{eqnarray}
\label{sec-3-1-pure-gauge-ansatz-der-coff}
c_{s,n}&=&(n+1)(2d+4s-4n)a_{s,n}+(s-2n+2)(s-2n+1)a_{s,n-1}\\
&+&4(n+1)a_{s-1,n}+2(s-2n+1)a_{s-1,n-1}+a_{s-2,n-1},\nonumber
\end{eqnarray}
with the constraints that $a_{s,n}=0$ for $n\geq[\frac{s}{2}]+1$ or $s\leq{-1}$ or $n\leq{-1}$, which are apparently absent from Eq.~(\ref{sec-3-1-pure-gauge-ansatz}). In the above, $d$ is the dimension of the space-time, and hereafter we shall work with $d=4$. Therefore, a consistent solution of Eq.~(\ref{sec-3-1-pure-gauge-eq-re}) exists if
\begin{eqnarray}
\label{sec-3-1-pure-gauge-sol-0}
c_{0,0}&=&1,\hspace{2mm}(s=0)\\
\label{sec-3-1-pure-gauge-sol-1}
c_{s,n}&=&0.\hspace{2.5mm}(s\geq{1})
\end{eqnarray}
Eq.~(\ref{sec-3-1-pure-gauge-sol-1}) should be satisfied for $s\geq{1}$. It is a recursive identity, by which $a_{s,n}$ with larger $s$ and $n$ can be determined by $a_{s,n}$ with smaller $s$ and $n$. By this recursive identity, the first four equations for $a_{s,n}$ are
\begin{eqnarray}
\label{sec-3-1-pure-gauge-sol-00}
8a_{0,0}&=&1,\hspace{2mm}(s=0,n=0)\\
\label{sec-3-1-pure-gauge-sol-10}
12a_{1,0}+4a_{0,0}&=&0,\hspace{2mm}(s=1,n=0)\\
\label{sec-3-1-pure-gauge-sol-20}
16a_{2,0}+4a_{1,0}&=&0,\hspace{2mm}(s=2,n=0)\\
\label{sec-3-1-pure-gauge-sol-21}
24a_{2,1}+2a_{2,0}+2a_{1,0}+a_{0,0}&=&0.\hspace{2.5mm}(s=2,n=1)
\end{eqnarray}
Here we have used the relations that $a_{s,n}=0$ for $n\geq[\frac{s}{2}]+1$ or $s\leq{-1}$ or $n\leq{-1}$. For the equations derived from the recursive identity~(\ref{sec-3-1-pure-gauge-sol-1}), the unknown variables match the number of equations. So the solution of Eq.~(\ref{sec-3-1-pure-gauge-eq}) can be determined by the recursive equations such as Eqs.~(\ref{sec-3-1-pure-gauge-sol-00})-(\ref{sec-3-1-pure-gauge-sol-21}). The pure gauge field are hence obtained as
\begin{eqnarray}
\label{sec-3-1-pure-gauge-form}
A^{\mu}_{\mathrm{pure}}(x)=\partial^{\mu}\varphi(x).
\end{eqnarray}
From Eq.~(\ref{sec-3-1-pure-gauge-ansatz}), we see that $A^{\mu}_{\mathrm{pure}}(x)=0$ for the Lorentz gauge $\partial^{\mu}A_{\mu}(x)=0$. However, the pure gauge field determined by Eqs.~(\ref{sec-3-1-pure-gauge-ansatz}) and (\ref{sec-3-1-pure-gauge-form}) has a complicated formulation, and its transformation under gauge transformations is not transparent. In appendix~\ref{app:1}, we give arguments that it has the transformations as $A^{\mu}$ as we required. In appendix~\ref{app:2.2}, we present another solution of Eq.~(\ref{sec-3-1-pure-gauge-eq}) through a slightly different method, that is,
\begin{eqnarray}
\label{sec-3-1-app2-2-pure-gauge-form}
\hat{A}^{\mu}_{\mathrm{pure}}(x)&=&A^{\mu}(x)-\frac{1}{2}x_{\alpha}F^{\alpha\mu}(x)
+\frac{1}{6}x_{\alpha_1}x_{\alpha_2}\partial^{\alpha_1}F^{\alpha_2\mu}(x)\\
&+&\frac{1}{8}x^{\mu}x_{\alpha_1}\partial_{\alpha_2}F^{\alpha_1\alpha_2}(x)
+\frac{1}{16}x^{\rho}x_{\rho}\partial_{\alpha}F^{\mu\alpha}(x)+\cdots.\nonumber
\end{eqnarray}
By this formulation, $\hat{A}^{\mu}_{\mathrm{pure}}(x)$ apparently transforms as $A^{\mu}(x)$ under gauge transformations. Eq.~(\ref{sec-3-1-app2-2-pure-gauge-form}) is expected to be equivalent with Eq.~(\ref{sec-3-1-pure-gauge-form}) by similar arguments in appendix~\ref{app:1}, albeit a straightforward proof of their equivalence, which means that we attempt to derive Eq.~(\ref{sec-3-1-app2-2-pure-gauge-form}) from Eq.~(\ref{sec-3-1-pure-gauge-form}) by the method of simple decompositions and combinations, is not easy to implement.

\subsection{Non-Abelian Case}\label{sec:3.2}

For non-Abelian cases, we need to solve the equation
\begin{eqnarray}
\label{sec-3-2-pure-gauge-eq}
D^{\mu}(A_{\mu}-\frac{i}{g}U^{-1}\partial_{\mu}U)=0,\hspace{2mm}U(x)=\mathrm{exp}^{-ig\phi(x)},
\end{eqnarray}
where $\phi(x)=\phi^{a}(x)T^{a}$ taking values in the Lie algebra of $SU(N)$ and $D_{\mu}=\partial_{\mu}-ig[A_{\mu},\cdot\hspace{0.5mm}]$ stands for a covariant derivative. Expanded as a series, the pure gauge field is
\begin{eqnarray}
\label{sec-3-2-pure-gauge-exp}
A^{\mu}_{\mathrm{pure}}=\frac{i}{g}U^{-1}\partial^{\mu}U
=\partial^{\mu}\phi-\frac{i}{2}g[\partial^{\mu}\phi,\phi]-\frac{1}{6}g^2[[\partial^{\mu}\phi,\phi],\phi]+\cdots.
\end{eqnarray}
Because of  its non-linearity, Eq.~(\ref{sec-3-2-pure-gauge-eq}) is not easy to solve. We attempt to solve it by perturbative methods. We suppose Eq.~(\ref{sec-3-2-pure-gauge-eq}) has a series expansion as
\begin{eqnarray}
\label{sec-3-2-pure-gauge-eq-series}
\phi=\phi^{(0)}+g\phi^{(1)}+g^{2}\phi^{(2)}+\cdots.
\end{eqnarray}
Then Eq.~(\ref{sec-3-2-pure-gauge-eq}) can be solved by setting terms of the same power of the coupling $g$ to be zero. For the term independent of $g$, we obtain
\begin{eqnarray}
\label{sec-3-2-pure-gauge-eq-series-0}
\partial^{\mu}\partial_{\mu}\phi^{(0)}(x)=\mathscr{F}^{(0)},\hspace{2mm}\mathscr{F}^{(0)}=\partial^{\mu}A_{\mu}(x).
\end{eqnarray}
This equation has the same structure with Eq.~(\ref{sec-3-1-pure-gauge-eq-re}). So we can solve it by the ansatz (\ref{sec-3-1-pure-gauge-ansatz}). The solution is
\begin{eqnarray}
\label{sec-3-2-pure-gauge-eq-series-0-sol}
\phi^{(0)}(x)=\sum^{\infty}_{s=0}\sum^{[\frac{s}{2}]}_{n=0}a_{s,n}(x^{\rho}x_{\rho})^{n+1}
x^{\alpha_1}\cdots{x}^{\alpha_{s-2n}}\partial_{\alpha_1}\cdots\partial_{\alpha_{s-2n}}
(\partial^{\beta}\partial_{\beta})^{n}\mathscr{F}^{(0)},
\end{eqnarray}
with $\phi^{(0)}$ and $A_{\alpha}(x)$ taking values in the Lie algebra of $SU(N)$, and the coefficients $a_{s,n}$ solved by Eqs.~(\ref{sec-3-1-pure-gauge-sol-0}) and (\ref{sec-3-1-pure-gauge-sol-1}). For the term of $g$ of power one in Eq.~(\ref{sec-3-2-pure-gauge-eq}), we obtain
\begin{eqnarray}
\label{sec-3-2-pure-gauge-eq-series-1}
\partial^{\mu}\partial_{\mu}\phi^{(1)}&=&\mathscr{F}^{(1)},\\
\label{sec-3-2-pure-gauge-eq-series-1-def}
\frac{i}{2}[\partial_{\mu}\phi^{(0)},\phi^{(0)}]-i[A^{\mu},\partial_{\mu}\phi^{(0)}]&=&\mathscr{F}^{(1)}.
\end{eqnarray}
In Eq.~(\ref{sec-3-2-pure-gauge-eq-series-1-def})$, \mathscr{F}^{(1)}$ is known because $\phi^{(0)}$ has been given by Eq.~(\ref{sec-3-2-pure-gauge-eq-series-0-sol}). This equation also has the same structure with Eq.~(\ref{sec-3-1-pure-gauge-eq-re}). We can solve it by the ansatz (\ref{sec-3-1-pure-gauge-ansatz}) with $\mathcal{F}$ replaced by $\mathscr{F}^{(1)}$. The solution is
\begin{eqnarray}
\label{sec-3-2-pure-gauge-eq-series-1-sol}
\phi^{(1)}(x)=\sum^{\infty}_{s=0}\sum^{[\frac{s}{2}]}_{n=0}a_{s,n}(x^{\rho}x_{\rho})^{n+1}
x^{\alpha_1}\cdots{x}^{\alpha_{s-2n}}\partial_{\alpha_1}\cdots\partial_{\alpha_{s-2n}}
(\partial^{\beta}\partial_{\beta})^{n}\mathscr{F}^{(1)}.
\end{eqnarray}
For the coefficient of $g^{2}$ in Eq.~(\ref{sec-3-2-pure-gauge-eq}), we obtain
\begin{eqnarray}
\label{sec-3-2-pure-gauge-eq-series-2}
\mathscr{F}^{(2)}&=&\partial^{\mu}\partial_{\mu}\phi^{(2)},\\
\label{sec-3-2-pure-gauge-eq-series-2-def}
\mathscr{F}^{(2)}&=&\frac{i}{2}[\partial^{\mu}\phi^{(0)},\phi^{(1)}]+\frac{i}{2}[\partial^{\mu}\phi^{(1)},\phi^{(0)}]
+\frac{1}{6}[[\partial^{\mu}\phi^{(0)},\phi^{(0)}],\phi^{(0)}]\\
&-&i[A^{\mu},\partial_{\mu}\phi^{(1)}-\frac{i}{2}[\partial^{\mu}\phi^{(0)},\phi^{(0)}]].\nonumber
\end{eqnarray}
This equation can be solved as above. The solution is
\begin{eqnarray}
\label{sec-3-2-pure-gauge-eq-series-2-sol}
\phi^{(2)}(x)=\sum^{\infty}_{s=0}\sum^{[\frac{s}{2}]}_{n=0}a_{s,n}(x^{\rho}x_{\rho})^{n+1}
x^{\alpha_1}\cdots{x}^{\alpha_{s-2n}}\partial_{\alpha_1}\cdots\partial_{\alpha_{s-2n}}
(\partial^{\beta}\partial_{\beta})^{n}\mathscr{F}^{(2)}.
\end{eqnarray}
By the foregoing recursive procedures, a solution of Eq.~(\ref{sec-3-2-pure-gauge-eq}) can be derived for the non-Abelian case. The pure gauge field $A^{\mu}_{\mathrm{pure}}$ hence can be obtained for the non-Abelian case. It is apparent that $A^{\mu}_{\mathrm{pure}}(x)=0$ in the Lorentz gauge $\partial^{\mu}A_{\mu}(x)=0$ by the foregoing constructions. Another solution of Eq.~(\ref{sec-3-2-pure-gauge-eq}) can be derived by following the procedure in appendix~\ref{app:3}, which is expected to be a generalization of Eq.~(\ref{sec-3-1-app2-2-pure-gauge-form})
\begin{eqnarray}
\label{sec-3-2-app3-2-pure-gauge-form}
\hat{A}^{\mu}_{\mathrm{pure}}(x)&=&A^{\mu}(x)-\frac{1}{2}x_{\alpha}F^{\alpha\mu}(x)
+\frac{1}{6}x_{\alpha_1}x_{\alpha_2}D^{\alpha_1}F^{\alpha_2\mu}(x)\\
&+&\frac{1}{8}x^{\mu}x_{\alpha_1}D_{\alpha_2}F^{\alpha_1\alpha_2}(x)
+\frac{1}{16}x^{\rho}x_{\rho}D_{\alpha}F^{\mu\alpha}(x)+\cdots,\nonumber
\end{eqnarray}
which is also expected to be equivalent with the solution (\ref{sec-3-2-pure-gauge-exp}) obtained above, albeit a formal proof is not easy to implement.

So far, we have derived expressions for the pure gauge field in the Abelian case and the non-Abelian case. These expressions shall be used in the next section to construct gauge invariant operators of the gluon spin and the photon spin.

\section{Gauge Invariant Decompositions of the Nucleon Spin}\label{sec:4}

\subsection{Generation of Gluon Spin Angular Momentum from Surface Currents}\label{sec:4.1}

In section~\ref{sec:2}, we interpreted the gauge invariant current of angular momentum as summations of gauge variant ones. We saw that the gauge invariant decomposition of Ji and that of Chen et al. can be accommodated into this framework. In the decomposition of Chen et al.~\cite{Chen:2008ag}, there is gauge invariant description of gluon spin. However, in the decomposition of Ji~\cite{Ji:1996ek}, there is no gauge invariant description of gluon spin, while only the total gluon angular momentum current is gauge invariant. In this section, we attempt to propose another decomposition of the nucleon spin. We begin with Ji's decomposition, that is, the quark parts are
\begin{eqnarray}
\label{sec-4-1-noether-quark-spin}
M_{qs}^{\mu\alpha\beta}&=&\frac{1}{2}\epsilon^{\mu\alpha\beta\rho}\bar{\psi}\gamma_{\rho}\gamma_{5}\psi,\\
\label{sec-4-1-noether-quark-orbit}
M_{qo}^{\mu\alpha\beta}&=&\frac{i}{2}[\bar{\psi}\gamma^{\mu}x^{\alpha}(\partial^{\beta}-igA^{\beta})\psi-(\alpha\longleftrightarrow
\beta)]+\mathrm{H.C.}\\
&+&({\eta}^{\mu\alpha}x^{\beta}-{\eta}^{\mu\beta}x^{\alpha}){\mathcal{L}}_{\mathrm{quark}},\nonumber
\end{eqnarray}
and the gluon part is
\begin{eqnarray}
\label{sec-4-1-noether-gluon-orbit}
M_{g}^{\mu\alpha\beta}&=&2\mathrm{Tr}\left[F^{\mu}_{\hspace{2mm}\nu}(F^{\nu\alpha}x^{\beta}-F^{\nu\beta}x^{\alpha})\right]\\
&+&({\eta}^{\mu\alpha}x^{\beta}-{\eta}^{\mu\beta}x^{\alpha}){\mathcal{L}}_{\mathrm{gluon}}.\nonumber
\end{eqnarray}
Apparently there is no natural candidate of the gluon spin operator in the above decomposition. In order to introduce descriptions of the gluon spin, following the similar proposal of Belinfante for the energy-momentum tensor~\cite{Belinfante:1940q}, we construct a gauge invariant surface current
\begin{eqnarray}
\label{sec-4-1-noether-gluon-surface}
\mathcal{M}_{g}^{\mu\alpha\beta}&=&
\frac{1}{2}\partial_{\rho}(x^{\alpha}\mathscr{M}_{g}^{\mu\rho\beta}-x^{\beta}\mathscr{M}_{g}^{\mu\rho\alpha}),\\
\label{sec-4-1-noether-gluon-surface-def}
\mathscr{M}_{g}^{\mu\alpha\beta}&=&-2\mathrm{Tr}(F^{\mu\alpha}A^{\beta}_{\mathrm{phys}}+F^{\beta\mu}A^{\alpha}_{\mathrm{phys}}
+F^{\alpha\beta}A^{\mu}_{\mathrm{phys}}).
\end{eqnarray}
Here the physical field $A^{\mu}_{\mathrm{phys}}=A^{\mu}-A^{\mu}_{\mathrm{pure}}$ has been defined in section~\ref{sec:3}. Apparently $\mathcal{M}_{g}^{\mu\alpha\beta}$ is also a conserved current, that is,
\begin{eqnarray}
\label{sec-4-1-noether-gluon-surface-eq}
\partial_{\mu}\mathcal{M}_{g}^{\mu\alpha\beta}=0,
\end{eqnarray}
because $\mathscr{M}_{g}^{\mu\rho\beta}$ is antisymmetrical about its indices $\mu$ and $\rho$. Adding this current to $M^{\mu\alpha\beta}$, we can obtain a new conserved current
\begin{eqnarray}
\label{sec-4-1-noether-gluon-cur-new}
\tilde{M}^{\mu\alpha\beta}=\mathcal{M}_{g}^{\mu\alpha\beta}+M^{\mu\alpha\beta}.
\end{eqnarray}
The gluon parts of this new current are
\begin{eqnarray}
\label{sec-4-1-noether-gluon-spin-new}
\tilde{M}_{gs}^{\mu\alpha\beta}&=&-2\mathrm{Tr}(F^{\mu\alpha}A^{\beta}_{\mathrm{phys}}+F^{\beta\mu}A^{\alpha}_{\mathrm{phys}}
+F^{\alpha\beta}A^{\mu}_{\mathrm{phys}}),\\
\label{sec-4-1-noether-gluon-orbit-new}
\tilde{M}_{go}^{\mu\alpha\beta}&=&2\mathrm{Tr}[F^{\mu}_{\hspace{2mm}\nu}(F^{\nu\alpha}x^{\beta}-F^{\nu\beta}x^{\alpha})]
+\mathrm{Tr}[x^{\beta}F^{\mu\rho}D_{\rho}A^{\alpha}_{\mathrm{phys}}
-x^{\alpha}F^{\mu\rho}D_{\rho}A^{\beta}_{\mathrm{phys}}]\nonumber\\
&+&\mathrm{Tr}[x^{\beta}D_{\rho}F^{\mu\rho}A^{\alpha}_{\mathrm{phys}}
-x^{\alpha}D_{\rho}F^{\mu\rho}A^{\beta}_{\mathrm{phys}}]\\
&+&\mathrm{Tr}[x^{\alpha}A^{\rho}_{\mathrm{phys}}D_{\rho}F^{\mu\beta}
-x^{\beta}A^{\rho}_{\mathrm{phys}}D_{\rho}F^{\mu\alpha}]
+({\eta}^{\mu\alpha}x^{\beta}-{\eta}^{\mu\beta}x^{\alpha}){\mathcal{L}}_{\mathrm{gluon}}.\nonumber
\end{eqnarray}
Here $D_{\mu}=\partial_{\mu}-ig[A_{\mu},\cdot\hspace{0.5mm}]$ and we have used the relation $D_{\rho}A^{\rho}_{\mathrm{phys}}=0$ as it is the definition of $A^{\rho}_{\mathrm{phys}}$ in section~\ref{sec:3.2}. The quark parts of this new current remain the same as Eqs.~(\ref{sec-4-1-noether-quark-spin}) and (\ref{sec-4-1-noether-quark-orbit}). The above gluon spin operator~(\ref{sec-4-1-noether-gluon-spin-new}) has been discussed in~\cite{Zhang:2011rn} in the light-cone gauge. The second line of Eq.~(\ref{sec-4-1-noether-gluon-orbit-new}) is reminiscent of the potential angular momentum in~\cite{Wakamatsu:2010cb}. Of course, the surface current (\ref{sec-4-1-noether-gluon-surface}) is not the unique current which can be used to construct descriptions of the gluon spin. For other choice, we can consider the current
\begin{eqnarray}
\label{sec-4-1-noether-gluon-surface-def-2}
\mathscr{M}_{g}^{\mu\alpha\beta}&=&-2\mathrm{Tr}(F^{\mu\alpha}A^{\beta}_{\mathrm{phys}}+F^{\beta\mu}A^{\alpha}_{\mathrm{phys}}
+F^{\alpha\beta}A^{\mu}_{\mathrm{phys}})
-2ig\mathrm{Tr}(A^{\mu}_{\mathrm{phys}}[A^{\alpha}_{\mathrm{phys}},A^{\beta}_{\mathrm{phys}}]).
\end{eqnarray}
This choice induces the gauge invariant Chern-Simons current as descriptions of the gluon spin. We shall make more discussions of appropriate descriptions of the gluon spin in the subsequent two subsections.

\subsection{Analysis on Laguerre-Gauss Laser Modes}\label{sec:4.2}

In this subsection, when restricted on the Abelian case, we discuss how the operator~(\ref{sec-4-1-noether-gluon-spin-new}) can be an appropriate description of the photon spin with a fixed frequency, such as the Laguerre-Gauss Laser Modes. As shown by Allen et al.~\cite{Allen:1992zz,Barnett:94B}, the Laguerre-Gaussian laser modes have a well-defined angular momentum, and the angular momentum can be well described by classical analysis without the need of quantum field theories. The Laguerre-Gaussian laser modes are optical modes with a fixed frequency, and they conform to the source-free Maxwell equations
\begin{eqnarray}
\label{sec-4-2-maxwell-eq}
\nabla\times \vec{E}=-\frac{\partial \vec{B}}{\partial{t}}, \hspace{2mm}\nabla\times \vec{B}=\frac{\partial \vec{E}}{\partial{t}}.
\end{eqnarray}
For $\vec{E}$ and $\vec{B}$ with the time dependence $e^{-i\omega{t}}$, the above equations can be rewritten as
\begin{eqnarray}
\label{sec-4-2-maxwell-eq-exp}
\vec{B}=-\frac{i}{\omega}\nabla\times \vec{E}, \hspace{2mm}\vec{E}=\frac{i}{\omega}\nabla\times \vec{B}.
\end{eqnarray}
In the above descriptions, only gauge invariant field strength are needed. Of course, we can also describe the Laguerre-Gaussian modes by using the gauge potential $A^{\mu}=(A^{0},\vec{A})$. Then the field strength $\vec{E}$ and $\vec{B}$ can be expressed by
\begin{eqnarray}
\label{sec-4-2-maxwell-pot}
\vec{E}=-\nabla{A}^{0}-\frac{\partial \vec{A}}{\partial{t}},\hspace{2mm}\vec{B}=\nabla\times \vec{A}.
\end{eqnarray}
To solve the Maxwell equations, we have the freedom to choose a convenient gauge. We consider the Lorentz gauge
\begin{eqnarray}
\label{sec-4-2-maxwell-lz-gau}
\partial^{\mu}A_{\mu}=\frac{\partial {A}^{0}}{\partial{t}}+\nabla\cdot\vec{A}=0.
\end{eqnarray}
In the Lorentz gauge, for optical modes with a fixed frequency, ${A}^{0}$ and $\vec{A}$ can both have the time dependence $e^{-i\omega{t}}$, which are consistent with the Lorentz gauge condition~(\ref{sec-4-2-maxwell-lz-gau}). So Eq.~(\ref{sec-4-2-maxwell-pot}) can be rewritten as
\begin{eqnarray}
\label{sec-4-2-maxwell-pot-re}
\vec{E}=-\nabla{A}^{0}+i\omega\vec{A},\hspace{2mm}\vec{B}=\nabla\times \vec{A}.
\end{eqnarray}

Now we begin to consider the descriptions of the photon spin. In the Lorentz gauge $\partial^{\mu}A_{\mu}=0$, we know that $A^{\mu}_{\mathrm{pure}}=0$ from the discussions in section~\ref{sec:3.1}. So the photon spin operator~(\ref{sec-4-1-noether-gluon-spin-new}) in the Lorentz gauge is simplified to be
\begin{eqnarray}
\label{sec-4-2-noether-gluon-spin}
\tilde{M}_{gs}^{\mu\alpha\beta}=-(F^{\mu\alpha}A^{\beta}+F^{\beta\mu}A^{\alpha}
+F^{\alpha\beta}A^{\mu}).
\end{eqnarray}
We define the photon spin vector as
\begin{eqnarray}
\label{sec-4-2-noether-gluon-spin-vec-def}
S^{i}&=&\frac{1}{2}\epsilon^{ijk}\int{d^{3}x}\tilde{M}_{gs}^{0jk},
\end{eqnarray}
when written manifestly, which is,
\begin{eqnarray}
\label{sec-4-2-noether-gluon-spin-vec}
\vec{S}&=&\int{d^{3}x}(\vec{E}\times\vec{A}+\vec{B}{A}^{0}).
\end{eqnarray}
Because $\vec{E}$, $\vec{B}$, ${A}^{0}$ and $\vec{A}$ all have the time dependence $e^{-i\omega{t}}$, as known in electrodynamics, the average of the photon spin vector over a time period $\frac{2\pi}{\omega}$ can be given by
\begin{eqnarray}
\label{sec-4-2-noether-gluon-spin-vec-ave}
\langle\vec{S}\rangle=\frac{1}{2}\int{d^{3}x}(\vec{E}^{\ast}\times\vec{A}+\vec{E}\times\vec{A}^{\ast}
+\vec{B}^{\ast}{A}^{0}+\vec{B}{A}^{0\ast}).
\end{eqnarray}
Here $Z^{\ast}$ means the complex conjugate of $Z$. Using the relation of $\vec{E}$ and $\vec{B}$ in Eq.~(\ref{sec-4-2-maxwell-eq-exp}), we obtain
\begin{eqnarray}
\label{sec-4-2-noether-gluon-spin-vec-ave-exp}
\langle\vec{S}\rangle=\frac{1}{2}\int{d^{3}x}(\vec{E}^{\ast}\times\vec{A}+\vec{E}\times\vec{A}^{\ast}
+\frac{i}{\omega}\nabla\times\vec{E}^{\ast}{A}^{0}-\frac{i}{\omega}\nabla\times\vec{E}{A}^{0\ast}).
\end{eqnarray}
After integrating by parts and supposing the surface terms to be zero, we obtain
\begin{eqnarray}
\label{sec-4-2-noether-gluon-spin-vec-ave-ip}
\langle\vec{S}\rangle=\frac{1}{2}\int{d^{3}x}[\vec{E}^{\ast}\times(\vec{A}+\frac{i}{\omega}\nabla{A}^{0})
+\vec{E}\times(\vec{A}^{\ast}-\frac{i}{\omega}\nabla{A}^{0\ast})].
\end{eqnarray}
Then using the definition of $\vec{E}$ in Eq.~(\ref{sec-4-2-maxwell-pot-re}), finally we obtain
\begin{eqnarray}
\label{sec-4-2-noether-gluon-spin-vec-ave-fin}
\langle\vec{S}\rangle=-\frac{i}{\omega}\int{d^{3}x}(\vec{E}^{\ast}\times\vec{E}).
\end{eqnarray}
This is consistent with the results in~\cite{Barnett:94B,Li:09a,Ji:2012gc}.

We give some comments here. For an Abelian gauge theory, the conventional Chern-Simons current~(\ref{sec-4-2-noether-gluon-spin}) appears to be the reduced version of the extended Chern-Simons current~(\ref{sec-4-1-noether-gluon-surface-def-2}) after the Lorentz gauge fixing $\partial^{\mu}A_{\mu}(x)=0$.  We just show that the Chern-Simons current works well as descriptions of the spin of optical modes with a fixed frequency. From the above discussions, we may get the impression that the conventional Chern-Simons current~(\ref{sec-4-2-noether-gluon-spin}) is sufficient for descriptions of the photon spin. However, this impression is not right. Although the Chern-Simons current ~(\ref{sec-4-2-noether-gluon-spin}) works well in classical descriptions as above, it does not work as descriptions in quantum theories. As shown by Manohar in the two dimensional Schwinger model~\cite{Manohar:1990eu}, a forward matrix element of the Chern-Simons current is not gauge invariant. However, a gauge invariant extension of the Chern-Simons current~(\ref{sec-4-1-noether-gluon-surface-def-2}) can yield gauge invariant results as shown in~\cite{Guo:2012wv}. Therefore, it is the gauge invariant extension of the Chern-Simons current~(\ref{sec-4-1-noether-gluon-surface-def-2}) works well both in classical and quantum theories.

Besides, the foregoing analysis apparently only applies to Abelian gauge theories. Because of the nonlinearity of non-Abelian theories, we do not have an appropriate definition of gluon modes with a fixed frequency. Therefore, for non-Abelian theories, we still can not determine whether it is the operator (\ref{sec-4-1-noether-gluon-spin-new}) or the operator (\ref{sec-4-1-noether-gluon-surface-def-2}) that can be considered as appropriate descriptions of the gluon spin by the above analysis, because they degenerate into the same formulation when applied to an Abelian theory.

\subsection{Relations to Parton Distribution Functions}\label{sec:4.3}

As shown in~\cite{Ji:1996ek}, the quark orbital angular momentum~(OAM) can be connected to the generalized parton distributions. In this subsection, we discuss the relations between the operators constructed in section~\ref{sec:4.1} and the parton distribution functions. We consider the parton distribution function
\begin{eqnarray}
\label{sec-4-3-pd-quark}
f_{q}(\xi)=\frac{1}{2P^{+}}\int\frac{d\lambda}{2\pi}e^{i\lambda{\xi}}\langle{PS}\vert\bar{\psi}(0)\gamma^{+}
\mathcal {U}[0;\lambda{n}]\psi(\lambda{n})\vert{PS}\rangle.
\end{eqnarray}
Here $n^{\mu}=\frac{1}{\sqrt{2}}(1,0,0,1)$ is a light-like vector, and $\mathcal {U}[0;\lambda{n}]$ is a path-ordered gauge link
\begin{eqnarray}
\label{sec-4-3-link}
\mathcal {U}[0;\lambda{n}]=\mathcal{P}\exp\left(ig\int_{0}^{\lambda}n_{\mu}N^{\mu}(u{n})du\right),
\end{eqnarray}
which makes $f_{q}(\xi)$ gauge invariant. The quark momentum can be connected to the second moment of $f_{q}(\xi)$
\begin{eqnarray}
\label{sec-4-3-pd-quark-mom}
\int{d}\xi\hspace{0.5mm}{\xi}f_{q}(\xi)=n_{\mu_1}n_{\mu_2}\langle{PS}\vert\bar{\psi}(0)\gamma^{\mu_1}
i(\partial^{\mu_2}-igN^{\mu_2})\psi(0)\vert{PS}\rangle.
\end{eqnarray}
For $N^{\mu}=A^{\mu}$, we obtain the gauge invariant quark momentum of Ji~\cite{Ji:1996ek}; For $N^{\mu}=A^{\mu}_{\mathrm{pure}}$, we obtain the quark momentum considered by Chen et al.~\cite{Chen:2009mr}. The above definition of parton distributions includes the gauge link $\mathcal {U}[0;\lambda{n}]$, which depends on the choice of path~(For a discussion, see~\cite{Wakamatsu:2013voa}). However, we can use the Dirac variables~\cite{Dirac:1955uv,Pervushin:2001kq,Lorce:2013gxa} to formulate another definition of parton distributions, which are free of the path dependence. In section~\ref{sec:3}, in order to derive expressions of $A^{\mu}_{\mathrm{pure}}$, we have derived manifest expressions of the unitary matrix $U(x)$, which can be used to formulate the Dirac variable
\begin{eqnarray}
\label{sec-4-3-quark-dirac}
\psi_{D}(x)=U(x)\psi(x).
\end{eqnarray}
By the definition of $U(x)$ in section~\ref{sec:3}, $\psi_{D}(x)$ is apparently gauge invariant. The parton distributions can be constructed from Dirac variables as
\begin{eqnarray}
\label{sec-4-3-pd-quark-dirac}
f^{D}_{q}(\xi)=\frac{1}{2P^{+}}\int\frac{d\lambda}{2\pi}e^{i\lambda{\xi}}\langle{PS}\vert\bar{\psi}(0)U^{-1}(0)\gamma^{+}
U(\lambda{n})\psi(\lambda{n})\vert{PS}\rangle,
\end{eqnarray}
which has the bi-local formulation $\psi_{D}(x)\psi_{D}(0)$, and each term of this bi-local expression is separately gauge invariant. Its second moment gives
\begin{eqnarray}
\label{sec-4-3-pd-quark-dirac-mom}
\int{d}\xi\hspace{0.5mm}{\xi}f^{D}_{q}(\xi)=n_{\mu_1}n_{\mu_2}\langle{PS}\vert\bar{\psi}(0)\gamma^{\mu_1}
i(\partial^{\mu_2}-ig\cdot\frac{i}{g}U^{-1}\partial^{\mu_2}U)\psi(0)\vert{PS}\rangle.
\end{eqnarray}
It shows that the pure field $A^{\mu}_{\mathrm{pure}}=\frac{i}{g}U^{-1}\partial^{\mu}U$ appears in the covariant derivative. So the gauge invariant extension of the canonical quark momentum based on the pure gauge field is closely connected to the parton distribution based on Dirac variables. From the definition of $A^{\mu}_{\mathrm{pure}}$ in section~\ref{sec:3}, we saw that the quark momentum of Chen et al. can be obtained by subtracting  a series operator contributions from that definition of Ji.

When the above applied to the gluon distributions, we have the gluon distributions based on Dirac variables
\begin{eqnarray}
\label{sec-4-3-pd-gluon}
f^{D}_{g}(\xi)=\frac{i}{2\xi{P}^{+}}\int\frac{d\lambda}{2\pi}e^{i\lambda{\xi}}\langle{PS}\vert
{U}(0){F}^{+\alpha}(0)U^{-1}(0)
{U}(\lambda{n}){\tilde{F}}^{\hspace{1mm}+}_{\alpha}(\lambda{n})U^{-1}(\lambda{n})\vert{PS}\rangle.
\end{eqnarray}
Here the Dirac variable ${U}(x){F}^{\mu\nu}(x)U^{-1}(x)$ is gauge invariant by the definition of ${U}(x)$ in section~\ref{sec:3}. The first moment of $f^{D}_{g}(\xi)$ is
\begin{eqnarray}
\label{sec-4-3-pd-gluon-dirac-mom}
\Delta{G}=\frac{1}{{P}^{+}}\int{d}\xi\hspace{0.5mm}f^{D}_{g}(\xi)=\frac{1}{{P}^{+}}\langle{PS}\vert{K}^{+}\vert{PS}\rangle.
\end{eqnarray}
Here ${K}^{\mu}$ is the Chern-Simons current which satisfies $\partial_{\mu}K^{\mu}=\frac{1}{2}F^{a}_{\mu\nu}\tilde{F}_{a}^{\mu\nu}$~\cite{Manohar:1990kr}. Because $f^{D}_{g}(\xi)$ is gauge invariant, we can regard $K^{\mu}$ as the gauge invariant extension of the Chern-Simons current. As shown in~\cite{Guo:2012wv}, the gauge invariant extension of the Chern-Simons current also satisfies $\partial_{\mu}K^{\mu}=\frac{1}{2}F^{a}_{\mu\nu}\tilde{F}_{a}^{\mu\nu}$. From this regard, it appears that the gauge invariant extension of the Chern-Simons current works as an appropriate description of the gluon spin.

\section{Conclusions}\label{sec:5}

We have discussed the feasibility of gauge invariant decompositions of the angular momentum of the gauge field. We examined the structure of the angular momentum current through the Noether theorem. The gauge invariant angular momentum currents are shown to be summations of gauge variant currents, which are conserved Noether currents induced separately by the Lorentz transformation and the gauge transformation. This summation formulation suggests that a gauge invariant measurement should include not only the conventional canonical current but also the gauge current. We construct novel expressions of the pure gauge field by using a tower of operators similar to twist operators in the operator product expansions. These expressions show that the pure gauge field includes contributions from an infinite operator series. It does not seem easy to sum this tower of operators when performing phenomenological calculations. We also discussed the appropriate operator description of light modes with a fixed frequency. The gauge invariant extension of the Chern-Simons current is shown to yield results consistent with other classical analysis. Regarding its frame-independence and its gauge-independence, the gauge-invariant Chern-Simons current could be a proper candidate for descriptions of the gluon spin, albeit its physical meaning is not transparent because of its manifest non-local formulation.

\acknowledgments

This work was supported in part by Fondecyt~(Chile) grant 1100287 and by Project Basal under Contract No.~FB0821.

\appendix

\section{Transformation of the Pure gauge Field}\label{app:1}

In this appendix, we give arguments that the pure gauge field ${A}^{\mu}_{\mathrm{pure}}$ derived in section~\ref{sec:3.1} transforms as ${A}^{\mu}$ under gauge transformations. For a gauge transformation, ${A}^{\mu}$ transforms as
\begin{eqnarray}
\label{app-1-gauge-tr}
\delta{A}_{\mu}(x)=\partial_{\mu}\Lambda(x).
\end{eqnarray}
Then by Eq.~(\ref{sec-3-1-pure-gauge-ansatz}), $\varphi$ transforms as
\begin{eqnarray}
\label{app-1-pure-gauge-tr}
\delta\varphi(x)&=&\sum^{\infty}_{s=0}\sum^{[\frac{s}{2}]}_{n=0}a_{s,n}(x^{\rho}x_{\rho})^{n+1}
x^{\alpha_1}\cdots{x}^{\alpha_{s-2n}}\partial_{\alpha_1}\cdots\partial_{\alpha_{s-2n}}
(\partial^{\beta}\partial_{\beta})^{n}\delta\mathcal{F}(x),\\
\label{app-1-pure-gauge-tr-def}
\delta\mathcal{F}&=&\partial_{\mu}\delta{A}^{\mu}=\partial_{\mu}\partial^{\mu}\Lambda(x).
\end{eqnarray}
From the discussions in section \ref{sec:3.1}, it is easy to understand that $\delta\varphi$ in Eq.~(\ref{app-1-pure-gauge-tr}) satisfies the equation
\begin{eqnarray}
\label{app-1-pure-gauge-tr-eq}
\partial_{\mu}\partial^{\mu}\delta\varphi(x)=\partial_{\mu}\partial^{\mu}\Lambda(x).
\end{eqnarray}
So $\delta\varphi$ can be solved as
\begin{eqnarray}
\label{app-1-pure-gauge-tr-eq-sol}
\delta\varphi(x)=h(x)+\Lambda(x), \hspace{2mm} \partial_{\mu}\partial^{\mu}h(x)=0.
\end{eqnarray}
Here $h(x)$ is independent of $\Lambda(x)$. So $\delta\varphi(x)$ equals to $\Lambda(x)$ up to a harmonic function. Moreover, from Eq.~(\ref{app-1-pure-gauge-tr}), we know that
\begin{eqnarray}
\label{app-1-pure-gauge-tr-eq-sol-1}
\Lambda(x)=0\Longrightarrow \delta\varphi(x)=0,
\end{eqnarray}
which means that $h(x)$ should be zero and actually $\delta\varphi(x)=\Lambda(x)$. So the pure gauge field ${A}^{\mu}_{\mathrm{pure}}=\partial^{\mu}\varphi(x)$ transforms as ${A}^{\mu}$.

\section{Another expression of the Pure Gauge Field: Abelian Case}\label{app:2}

In this appendix, we are going to derive another expression for the pure gauge field, which satisfies Eq.~(\ref{sec-3-1-pure-gauge-eq}). This expression looks different from Eq.~(\ref{sec-3-1-pure-gauge-ansatz}), but is expected to be equivalent to Eq.~(\ref{sec-3-1-pure-gauge-ansatz}). In order to implement this goal, we first solve the pure gauge field through the Fock-Schwinger gauge condition, which then can be used to find another solution of Eq.~(\ref{sec-3-1-pure-gauge-eq}).

\subsection{Fock-Schwinger Gauge in Abelian Case}\label{app:2.1}

For an Abelian gauge field, we consider a pure gauge field determined by the Fock-Schwinger gauge condition
\begin{eqnarray}
\label{app-2-1-pure-gauge-eq}
x^{\mu}[A_{\mu}(x)-\partial_{\mu}\varphi_{FS}(x)]=0.
\end{eqnarray}
To solve Eq.~(\ref{app-2-1-pure-gauge-eq}), we make the ansatz
\begin{eqnarray}
\label{app-2-1-pure-gauge-ansatz}
\varphi_{FS}(x)=b_{1}x^{\alpha}A_{\alpha}(x)
+\sum^{\infty}_{n=2}b_{n}x^{\alpha_1}\cdots{x}^{\alpha_n}\partial_{\alpha_1}\cdots\partial_{\alpha_{n-1}}A_{\alpha_n}(x).
\end{eqnarray}
This expansion is similar to but different from that in section~\ref{sec:3.1}. A consistent solution of Eq.~(\ref{app-2-1-pure-gauge-eq}) exists if
\begin{eqnarray}
\label{app-2-1-pure-gauge-sol}
b_{1}=1,\hspace{2mm}b_{2}=-\frac{1}{2},\hspace{2mm}b_{n}=-\frac{1}{n}b_{n-1},\hspace{2mm}n=3,4,5\cdots.
\end{eqnarray}
The solution Eq.~(\ref{app-2-1-pure-gauge-ansatz}) is not very concise at first sight. However, compared to those expressions in section~\ref{sec:3.1}, Eq.~(\ref{app-2-1-pure-gauge-ansatz}) has the nice property that, through some simple combinations, the pure gauge field that determined by it can be concisely expressed as
\begin{eqnarray}
\label{app-2-1-pure-gauge-pure}
\mathcal{A}^{\mu}_{\mathrm{pure}}&=&\partial^{\mu}\varphi_{FS}(x)\\
&=&A^{\mu}(x)-\frac{1}{2}x_{\alpha}F^{\alpha\mu}(x)
+\sum_{n=2}^{\infty}\frac{(-1)^{n}}{(n+1)!}x_{\alpha_1}\cdots{x}_{\alpha_n}
\partial^{\alpha_1}\cdots\partial^{\alpha_{n-1}}F^{\alpha_n\mu}(x),\nonumber
\end{eqnarray}
where $F^{\alpha\mu}(x)=\partial^{\alpha}A^{\mu}(x)-\partial^{\mu}A^{\alpha}(x)$ is the Abelian field strength, and then the corresponding physical field can be given as
\begin{eqnarray}
\label{app-2-1-pure-gauge-phys}
\mathcal{A}^{\mu}_{\mathrm{phys}}&=&A^{\mu}-\mathcal{A}^{\mu}_{\mathrm{pure}}\\
&=&\frac{1}{2}x_{\alpha}F^{\alpha\mu}(x)+\sum_{n=2}^{\infty}\frac{(-1)^{n+1}}{(n+1)!}x_{\alpha_1}\cdots{x}_{\alpha_n}
\partial^{\alpha_1}\cdots\partial^{\alpha_{n-1}}F^{\alpha_n\mu}(x),\nonumber
\end{eqnarray}
which satisfies the Fock-Schwinger gauge condition
\begin{eqnarray}
\label{app-2-1-fs}
x_{\mu}\mathcal{A}^{\mu}_{\mathrm{phys}}=0.
\end{eqnarray}
Similar but slightly different expressions of the physical field have been given in~\cite{Shifman:1980a,Novikov:1984wz} through different methods. From the above, we saw that the pure gauge field~(\ref{app-2-1-pure-gauge-pure}) that determined by the Fock-Schwinger gauge condition has a very concise expression and its transformation under gauge transformations is also transparent. These concise expressions can be used in next subsection to obtain concise expressions of the pure gauge field determined by the Lorentz gauge condition.

\subsection{Expression of the Pure Gauge Field in Abelian Case}\label{app:2.2}

Now we try to solve Eq.~(\ref{sec-3-1-pure-gauge-eq}) in another way. We decompose $\varphi(x)$ in Eq.~(\ref{sec-3-1-pure-gauge-eq}) as
\begin{eqnarray}
\label{app-2-2-decom}
\varphi(x)=\varphi_{FS}(x)+\tilde{\varphi}(x).
\end{eqnarray}
Then by Eqs.~(\ref{app-2-1-pure-gauge-pure}) and (\ref{app-2-1-pure-gauge-phys}), Eq.~(\ref{sec-3-1-pure-gauge-eq}) can be rewritten as
\begin{eqnarray}
\label{app-2-2-pure}
\partial^{\mu}\partial_{\mu}\tilde{\varphi}(x)=\tilde{\mathcal{F}},\hspace{2mm}\tilde{\mathcal{F}}
=\partial_{\mu}\mathcal{A}^{\mu}_{\mathrm{phys}}.
\end{eqnarray}
Here $\mathcal{A}^{\mu}_{\mathrm{phys}}$ is given by Eq.~(\ref{app-2-1-pure-gauge-phys}). This equation has the same structure with Eq.~(\ref{sec-3-1-pure-gauge-eq-re}). It can be solved by the ansatz Eq.~(\ref{sec-3-1-pure-gauge-ansatz}), that is,
\begin{eqnarray}
\label{app2-2-pure-gauge-ansatz}
\tilde{\varphi}(x)=\sum^{\infty}_{s=0}\sum^{[\frac{s}{2}]}_{n=0}a_{s,n}(x^{\rho}x_{\rho})^{n+1}
x^{\alpha_1}\cdots{x}^{\alpha_{s-2n}}\partial_{\alpha_1}\cdots\partial_{\alpha_{s-2n}}
(\partial^{\beta}\partial_{\beta})^{n}\tilde{\mathcal{F}}.
\end{eqnarray}
In the Abelian case, $\tilde{\mathcal{F}}$ is apparently gauge invariant by its definition in Eq.~(\ref{app-2-2-pure}) and by Eq.~(\ref{app-2-1-pure-gauge-phys}). So $\tilde{\varphi}(x)$ is also gauge invariant. By the above solutions, ${A}^{\mu}_{\mathrm{pure}}$ can be written as
\begin{eqnarray}
\label{app2-2-pure-gauge-form}
A^{\mu}_{\mathrm{pure}}(x)&=&\partial^{\mu}\varphi
=\partial^{\mu}\varphi_{FS}+\partial^{\mu}\tilde{\varphi}=\mathcal{A}^{\mu}_{\mathrm{pure}}+\tilde{A}^{\mu}_{\mathrm{pure}}\nonumber\\
&=&A^{\mu}(x)-\frac{1}{2}x_{\alpha}F^{\alpha\mu}(x)+\frac{1}{6}x_{\alpha_1}x_{\alpha_2}\partial^{\alpha_1}F^{\alpha_2\mu}(x)\\
&+&\frac{1}{8}x^{\mu}x_{\alpha_1}\partial_{\alpha_2}F^{\alpha_1\alpha_2}(x)
+\frac{1}{16}x^{\rho}x_{\rho}\partial_{\alpha}F^{\mu\alpha}(x)+\cdots.\nonumber
\end{eqnarray}
From the above, we saw that the pure gauge field is decomposed into two parts: $\mathcal{A}^{\mu}_{\mathrm{pure}}$ and $\tilde{A}^{\mu}_{\mathrm{pure}}$. The part $\mathcal{A}^{\mu}_{\mathrm{pure}}$ is determined by the Fock-Schwinger gauge condition, and it has a very concise expression~(\ref{app-2-1-pure-gauge-pure}). The part $\tilde{A}^{\mu}_{\mathrm{pure}}$ is determined by Eq.~(\ref{app2-2-pure-gauge-ansatz}), which depends on $\mathcal{A}^{\mu}_{\mathrm{phys}}$ in Eq.~(\ref{app-2-1-pure-gauge-phys}). That is the above decomposition makes the solution~(\ref{app2-2-pure-gauge-form}) to be possible. Here we only write out the terms of two derivatives, albeit a expression of all orders is also manifest by the above constructions. This new formulation makes the gauge transformation of the pure gauge field transparent.

\section{Another expression of the Pure Gauge Field: Non-Abelian Case}\label{app:3}

\subsection{Fock-Schwinger Gauge in Non-Abelian Case}\label{app:3.1}

Similar to the discussion of Fock-Schwinger gauge condition in the Abelian case, for non-Abelian cases, we need to solve the equation
\begin{eqnarray}
\label{app-3-1-pure-gauge-eq}
x^{\mu}[A_{\mu}-\frac{i}{g}U_{FS}^{-1}\partial_{\mu}U_{FS}]=0,\hspace{2mm}U_{FS}(x)=\mathrm{exp}^{-ig\phi_{FS}(x)},
\end{eqnarray}
where $\phi(x)=\phi^{a}(x)T^{a}$ taking values in the Lie algebra of $SU(N)$. Expanded as a series, the pure gauge field is
\begin{eqnarray}
\label{app-3-1-pure-gauge-exp}
\mathcal{A}^{\mu}_{\mathrm{pure}}=\frac{i}{g}U_{FS}^{-1}\partial^{\mu}U_{FS}
=\partial^{\mu}\phi_{FS}-\frac{i}{2}g[\partial^{\mu}\phi_{FS},\phi_{FS}]-\frac{1}{6}g^2[[\partial^{\mu}\phi_{FS},\phi_{FS}],\phi_{FS}]+\cdots.
\end{eqnarray}
Because of its non-linearity, we attempt to solve Eq.~(\ref{app-3-1-pure-gauge-eq}) perturbatively. We suppose Eq.~(\ref{app-3-1-pure-gauge-eq}) has a series solution
\begin{eqnarray}
\label{app-3-1-pure-gauge-eq-series}
\phi_{FS}=\phi_{FS}^{(0)}+g\phi_{FS}^{(1)}+g^{2}\phi_{FS}^{(2)}+\cdots.
\end{eqnarray}
Then Eq.~(\ref{app-3-1-pure-gauge-eq}) can be solved by setting terms of the same power of the coupling $g$ to be zero. For the term independent of $g$, we obtain
\begin{eqnarray}
\label{app-3-1-pure-gauge-eq-series-0}
x^{\mu}[A_{\mu}-\partial_{\mu}\phi_{FS}^{(0)}]=0.
\end{eqnarray}
This equation has the same structure with Eq.~(\ref{app-2-1-pure-gauge-eq}). So we can solve it by the ansatz (\ref{app-2-1-pure-gauge-ansatz}). The solution is
\begin{eqnarray}
\label{app-3-1-pure-gauge-eq-series-0-sol}
\phi_{FS}^{(0)}(x)=b_{1}x^{\alpha}A_{\alpha}(x)
+\sum^{\infty}_{n=2}b_{n}x^{\alpha_1}\cdots{x}^{\alpha_n}\partial_{\alpha_1}\cdots\partial_{\alpha_{n-1}}A_{\alpha_n}(x),
\end{eqnarray}
with $\phi_{FS}^{(0)}$ and $A_{\alpha}(x)$ taking values in the Lie algebra of $SU(N)$, and the coefficients $b_{n}$ given by Eq.~(\ref{app-2-1-pure-gauge-sol}). For the term of $g$ of power one in Eq.~(\ref{app-3-1-pure-gauge-eq}), we obtain
\begin{eqnarray}
\label{app-3-1-pure-gauge-eq-series-1}
x^{\mu}[\mathscr{A}^{(1)}_{\mu}-\partial_{\mu}\phi_{FS}^{(1)}]&=&0,\\
\label{app-3-1-pure-gauge-eq-series-1-def}
\frac{i}{2}[\partial^{\mu}\phi_{FS}^{(0)},\phi_{FS}^{(0)}]&=&\mathscr{A}^{(1)}_{\mu}.
\end{eqnarray}
$\mathscr{A}^{(1)}_{\mu}$ is known because $\phi_{FS}^{(0)}$ is given by Eq.~(\ref{app-3-1-pure-gauge-eq-series-0-sol}). This equation also has the same structure with Eq.~(\ref{app-2-1-pure-gauge-eq}). We can solve it by the ansatz (\ref{app-2-1-pure-gauge-ansatz}) with $A_{\mu}$ replaced by $\mathscr{A}^{(1)}_{\mu}$. The solution is
\begin{eqnarray}
\label{app-3-1-pure-gauge-eq-series-1-sol}
\phi_{FS}^{(1)}(x)=b_{1}x^{\alpha}\mathscr{A}^{(1)}_{\alpha}(x)
+\sum^{\infty}_{n=2}b_{n}x^{\alpha_1}\cdots{x}^{\alpha_n}\partial_{\alpha_1}\cdots\partial_{\alpha_{n-1}}
\mathscr{A}^{(1)}_{\alpha_n}(x).
\end{eqnarray}
For the coefficient of $g^{2}$ in Eq.~(\ref{app-3-1-pure-gauge-eq}), we obtain
\begin{eqnarray}
\label{app-3-1-pure-gauge-eq-series-2}
x^{\mu}[\mathscr{A}^{(2)}_{\mu}-\partial_{\mu}\phi_{FS}^{(2)}]&=&0,\\
\label{app-3-1-pure-gauge-eq-series-2-def}
\frac{i}{2}[\partial^{\mu}\phi_{FS}^{(0)},\phi_{FS}^{(1)}]+\frac{i}{2}[\partial^{\mu}\phi_{FS}^{(1)},\phi_{FS}^{(0)}]
+\frac{1}{6}[[\partial^{\mu}\phi_{FS}^{(0)},\phi_{FS}^{(0)}],\phi_{FS}^{(0)}]&=&\mathscr{A}^{(2)}_{\mu}.
\end{eqnarray}
This equation can be solved as above. The solution is
\begin{eqnarray}
\label{app-3-1-pure-gauge-eq-series-2-sol}
\phi_{FS}^{(2)}(x)=b_{1}x^{\alpha}\mathscr{A}^{(2)}_{\alpha}(x)
+\sum^{\infty}_{n=2}b_{n}x^{\alpha_1}\cdots{x}^{\alpha_n}\partial_{\alpha_1}\cdots\partial_{\alpha_{n-1}}
\mathscr{A}^{(2)}_{\alpha_n}(x).
\end{eqnarray}
By the foregoing recursive procedure, a solution of Eq.~(\ref{app-3-1-pure-gauge-eq}) can be derived for non-Abelian cases. We saw that the solution of $\phi_{FS}$ has a very complicated formulation. However, the expression for $\mathcal{A}^{\mu}_{\mathrm{pure}}$ and $\mathcal{A}^{\mu}_{\mathrm{phys}}$ are very concise. They are straightforward generalizations of Eqs.~(\ref{app-2-1-pure-gauge-pure}) and (\ref{app-2-1-pure-gauge-phys}), and we have
\begin{eqnarray}
\label{app-3-1-pure-gauge-pure}
\mathcal{A}^{\mu}_{\mathrm{pure}}&=&\frac{i}{g}U_{FS}^{-1}\partial_{\mu}U_{FS}
=A^{\mu}(x)-\frac{1}{2}x_{\alpha}F^{\alpha\mu}(x)+\frac{1}{6}x_{\alpha_1}x_{\alpha_2}D^{\alpha_1}F^{\alpha_2\mu}(x)+\cdots,\\
\label{app-3-1-pure-gauge-phys}
\mathcal{A}^{\mu}_{\mathrm{phys}}&=&A^{\mu}-\mathcal{A}^{\mu}_{\mathrm{pure}}
=\frac{1}{2}x_{\alpha}F^{\alpha\mu}(x)-\frac{1}{6}x_{\alpha_1}x_{\alpha_2}D^{\alpha_1}F^{\alpha_2\mu}(x)+\cdots,
\end{eqnarray}
where $F^{\alpha\mu}(x)=\partial^{\alpha}A^{\mu}(x)-\partial^{\mu}A^{\alpha}(x)-ig[A^{\alpha}(x),A^{\mu}(x)]$ is the non-Abelian field strength and $D^{\alpha}=\partial^{\alpha}-ig[A^{\alpha}(x),\cdot\hspace{0.5mm}]$. The physical field here also satisfies the Fock-Schwinger gauge condition.

\subsection{Expression of the Pure Gauge Field in Non-Abelian Case}\label{app:3.2}

In this appendix, we try to solve Eq.~(\ref{sec-3-2-pure-gauge-eq}) in another way, similar to the Abelian case, we decompose $\phi(x)$ as
\begin{eqnarray}
\label{app-3-2-decom}
\phi(x)=\phi_{FS}(x)+\tilde{\phi}(x).
\end{eqnarray}
In the series formulation, it is
\begin{eqnarray}
\label{app-3-2-decom-series}
\phi^{(0)}(x)&=&\phi^{(0)}_{FS}(x)+\tilde{\phi}^{(0)}(x),\phi^{(1)}(x)=\phi^{(1)}_{FS}(x)+\tilde{\phi}^{(1)}(x),\\
\phi^{(2)}(x)&=&\phi^{(2)}_{FS}(x)+\tilde{\phi}^{(2)}(x),\cdots.\nonumber
\end{eqnarray}
Then another solution of Eq.~(\ref{sec-3-2-pure-gauge-eq}) can be derived through the similar procedures in appendix~\ref{app:2.2} and section~\ref{sec:3.2}.

\bibliographystyle{utphys}
\bibliography{SpinGloun-Revisit-Ref}

\end{document}